\documentclass[a4paper,11pt]{article}
\usepackage{pos}

\usepackage{amsmath} 
\usepackage{graphicx}
\usepackage{dcolumn}
\usepackage{bm,hyperref}
\usepackage{float,amsfonts}
\usepackage{color}
\usepackage{tikz}
\usepackage{physics}
\usepackage{adjustbox}
\usetikzlibrary{decorations.pathreplacing}
\usetikzlibrary{arrows.meta}
\usepackage{subcaption}
\usepackage{wrapfig}

\newcommand{\SU}{\mathrm{SU}}
\newcommand{\Pf}{\mathcal{P}_{\mathrm{f}}}
\newcommand{\Pre}{\mathcal{P}_{\mathrm{r}}}
\newcommand{\Wd}{W_{\mathrm{d}}}
\newcommand{\DKL}{\tilde{D}_{\mathrm{KL}}}

\newcommand{\hESS}{\hat{\mathrm{ESS}}}
\newcommand{\nstep}{n_{\mathrm{step}}}
\newcommand{\ndof}{n_{\mathrm{dof}}}
\newcommand{\U}{\mathcal{U}}
\newcommand{\CP}{\mathrm{CP}}
\newcommand{\ee}{\mathrm{e}}

\title{A scalable flow-based approach to mitigate topological freezing}
\ShortTitle{A scalable flow-based approach to mitigate topological freezing}
\author[a,b]{Claudio Bonanno}
\author[c]{Andrea Bulgarelli}
\author*[d]{Elia Cellini}
\author[e]{Alessandro Nada}
\author[e]{Dario Panfalone}
\author[f]{Davide Vadacchino}
\author[e]{Lorenzo Verzichelli}

\affiliation[a]{Albert Einstein Center for Fundamental Physics, Institute for Theoretical Physics, University of Bern, Sidlerstra{\ss}e 5, CH-3012 Bern, Switzerland}
\affiliation[b]{Instituto de F\'isica Te\'orica UAM-CSIC, c/ Nicol\'as Cabrera 13-15, Universidad Aut\'onoma de Madrid, Cantoblanco, E-28049 Madrid, Spain}
\affiliation[c]{Transdisciplinary Research Area ``Building Blocks of Matter and Fundamental Interactions'' (TRA Matter) and Helmholtz Institute for Radiation and Nuclear Physics (HISKP), University of Bonn, Nussallee 14-16, 53115 Bonn, Germany}
\affiliation[d]{Higgs Centre for Theoretical Physics, School of Physics and Astronomy,
The University of Edinburgh, Edinburgh EH9 3FD, United Kingdom}
\affiliation[e]{Dipartimento di Fisica,  Universit\'a degli Studi di Torino and INFN, Sezione di Torino, Via Pietro Giuria 1, I-10125 Turin, Italy}
\affiliation[f]{Centre for Mathematical Sciences, University of Plymouth, Plymouth, PL4 8AA, United Kingdom}

\emailAdd{elia.cellini@ed.ac.uk}

\abstract{
As lattice gauge theories with non-trivial topological features are driven towards the continuum limit, standard Markov Chain Monte Carlo simulations suffer for topological freezing, i.e., a dramatic growth of autocorrelations in topological observables. A widely used strategy is the adoption of Open Boundary Conditions (OBC), which restores ergodic sampling of topology but at the price of breaking translation invariance and introducing unphysical boundary artifacts. In this contribution we summarize a scalable, exact flow-based strategy to remove them by transporting configurations from a prior with a OBC defect to a fully periodic ensemble, and apply it to $4d$ $\SU(3)$ Yang--Mills theory. The method is based on a Stochastic Normalizing Flow (SNF) that alternates non-equilibrium Monte Carlo updates with localized, gauge-equivariant defect coupling layers implemented via masked parametric stout smearing. Training is performed by minimizing the average dissipated work, equivalent to a Kullback--Leibler divergence between forward and reverse non-equilibrium path measures, to achieve more reversible trajectories and improved efficiency. We discuss the scaling with the number of degrees of freedom affected by the defect and show that defect SNFs achieve better performances than purely stochastic non-equilibrium methods at comparable cost. Finally, we validate the approach by reproducing reference results for the topological susceptibility.
}

\FullConference{The 42$^{\text{nd}}$ International Symposium on Lattice Field Theory (LATTICE2025)\\
2--8 November 2025\\
Tata Institute of Fundamental Research, Mumbai, India\\}

\begin{document}
\maketitle

\section{Introduction}

Standard lattice simulations of four-dimensional Yang--Mills theories sample gauge fields $ U $ from the target Boltzmann weight $ p(U)\propto \ee^{-S[U]} $ using Markov Chain Monte Carlo (MCMC) methods.
As the continuum limit is approached, autocorrelations grow because of critical slowing down, and the slowest modes are typically topological. This leads to a severe loss of ergodicity, known as topological freezing~\cite{Alles:1996vn,DelDebbio:2004xh,Schaefer:2010hu}.
A robust mechanism to accelerate topology sampling is the use of open boundary conditions (OBC) in time, which remove topological barriers and turn the evolution of topological modes into a diffusion process~\cite{Luscher:2011kk,Luscher:2012av,McGlynn:2014bxa}.
However, OBC break translation invariance and introduce unphysical boundary effects: the key algorithmic question is therefore whether it is possible to exploit the fast Monte Carlo dynamics of topological fluctuations of OBC while computing observables with periodic boundary conditions (PBC).

A successful answer to this problem is provided by Parallel Tempering on Boundary Conditions (PTBC), where replicas with different boundary conditions interpolating between OBC and PBC are simulated in parallel and allowed to swap configurations \cite{Hasenbusch:2017unr,Berni:2019bch,Bonanno:2022hmz,Bonanno:2020hht, Bonanno:2022yjr, Bonanno:2023hhp, Bonanno:2024nba, Bonanno:2024ggk,Bonanno:2024zyn,Bonanno:2025eeb}.
In this way, fast topological fluctuations generated with OBC are transferred to the physical PBC ensemble, while maintaining exactness.
Our approach~\cite{Bonanno:2025pdp} shares the same philosophy, but follows a different, flow-based route.

In this conference proceeding we focus on a flow-based strategy based on Stochastic Normalizing Flows (SNFs)~\cite{wu2020stochastic,Caselle:2022acb,Caselle:2024ent,Bulgarelli:2024brv,Bonanno:2025pdp,Kreit:2026eng}, a combination of Normalizing Flows~\cite{rezende2015variational,Albergo:2019eim} and Non-Equilibrium Markov Chain Monte Carlo (NE-MCMC) calculations based on Jarzynski and Crooks identities \cite{Jarzynski:1997ef,Crooks:1998,Caselle:2016wsw,Caselle:2018kap,Bulgarelli:2023ofi, Bulgarelli:2024onj,Bulgarelli:2025riv,Bonanno:2024udh}.
Crucially, SNFs preserve the favorable scaling of NE-MCMC with the number of degrees of freedom undergoing a transformation, while significantly reducing the dissipated work compared to purely stochastic protocols, translating into a substantial gain in efficiency. In particular, we exploit the NE-MCMC structure to design a controlled non-equilibrium interpolation between OBC and PBC, already explored in Ref.~\cite{Bonanno:2024udh} in $2d$ $\CP^{N-1}$ models and in Ref.~\cite{Bonanno:2025pdp} for $4d$ $\SU(3)$ Yang--Mills theory, that preserves exactness through reweighting and improves the sampling of topological sectors near the continuum limit.

More broadly, this work is part of a recent effort to go beyond purely local Monte Carlo updates by constructing learned transformations between ensembles in lattice field theory simulations. Normalizing flows have been widely investigated in lattice gauge theory~\cite{Kanwar:2020xzo, Boyda:2020hsi, Favoni:2020reg, Bacchio:2022, Abbott:2023thq, Gerdes:2024rjk,Albergo:2021bna, Finkenrath:2022ogg, Albergo:2022qfi, Abbott:2022zhs, Abbott:2024kfc} and realize exact, invertible maps between probability distributions, while related approaches based on diffusion models~\cite{Wang:2023exq, Zhu:2024kiu, Aarts:2024rsl, Zhu:2025pmw, Aarts:2025lpi,Vega:2025hgz,Kanwar:2025wuc}, generate configurations by reversing a stochastic noise process.

\section{Stochastic Normalizing Flows}
We consider the problem of transporting gauge configurations from a prior ensemble (that is easier to sample from) to a physical target one. Concretely, we introduce a prior distribution $ q_0(U_0)\propto \ee^{-S_0[U_0]} $ and a target distribution $ p(U)\propto \ee^{-S[U]} $ (here, OBC and PBC in $4d$ $\SU(3)$ gauge theory). The transport is implemented through a non-equilibrium evolution, defined as a sequence $ \U=[U_0,\dots,U_{\nstep}]\equiv[U_0,\dots,U] $ generated by a protocol $ \lambda(n) $ via Markov updates $P_{\lambda(n)} $ with equilibrium weight $\propto \ee^{-S_{\lambda(n)}}$.
A Jarzynski-based reweighting yields an unbiased estimator for target expectation values:
\begin{equation}
\langle \mathcal{O} \rangle_p
=
\frac{\langle \mathcal{O}(U)\, \ee^{-W(\U)} \rangle_{\mathrm{f}}}{\langle \ee^{-W(\U)} \rangle_{\mathrm{f}}}
\end{equation}
where $ \langle \cdot \rangle_{\mathrm{f}} $ is an average over forward trajectories and the work $W$ is
\begin{equation}
W(\U)=\sum_{n=0}^{\nstep-1}\Bigl(S_{\lambda(n+1)}[U_n]-S_{\lambda(n)}[U_n]\Bigr)\,.
\end{equation}
In practice, the exponential reweighting becomes inefficient if the dissipated work $ \Wd=W-\Delta F $, with $\Delta F$ being the free energy difference between prior and target, is large.

SNFs enhance these trajectories by inserting parametric, deterministic, invertible layers $ g_{\rho(n)} $ between stochastic updates:
\begin{equation}
U_0 \xrightarrow{\, g_{\rho(1)} \,} g_{\rho(1)}(U_0)
\xrightarrow{\, P_{\lambda(1)} \,} U_1
\xrightarrow{\, g_{\rho(2)} \,} \cdots
\xrightarrow{\, P_{\lambda(\nstep)} \,} U_{\nstep}\equiv U\,.
\end{equation}
The correct reweighting uses a variational work including Jacobians \cite{wu2020stochastic,Vaikuntanathan_2011,Caselle:2022acb}:
\begin{align}
W^{(\rho)}(\U)
&=
S[U]-S_0[U_0]-Q^{(\rho)}(\U)
-\sum_{n=0}^{\nstep-1}\log\left|\det J_{g_{\rho(n+1)}}(U_n)\right|\,,
\end{align}
where $ Q^{(\rho)}(\U) $ is a generalized pseudo-heat term (including the effect of inserting $ g_{\rho} $ layers).

A key quantity to control the performances of non-equilibrium samplers is the reverse Kullback--Leibler (KL) divergence between forward and reverse path probability densities:
\begin{equation}
\DKL\bigl(q_0\,\Pf \,\|\, p\,\Pre\bigr)=\langle \Wd \rangle_{\mathrm{f}} \ge 0\,,
\end{equation}
so making trajectories more reversible (smaller $ \langle \Wd \rangle $) directly stabilizes reweighting.
A commonly used empirical proxy for the efficiency of the reweighting estimator, expressed as an effective sample size, is
\begin{equation}
\hESS \equiv \frac{\langle \ee^{-W}\rangle_{\mathrm{f}}^2}{\langle \ee^{-2W}\rangle_{\mathrm{f}}}
= \frac{1}{\langle \ee^{-2\Wd}\rangle_{\mathrm{f}}}\,,
\end{equation}
showing that the effective sample size is controlled by fluctuations of the dissipated work, and that large positive values of $ \Wd $ exponentially suppress $ \hESS $.

\subsection{Defect gauge-equivariant coupling layers}

To construct efficient layers for gauge theories, we use gauge equivariant updates based on masked stout smearing \cite{Morningstar:2003gk,Nagai:2021bhh,Abbott:2023thq}.
At layer $ n $, a link is updated as
\begin{equation}
U'_\mu(x)=\exp\,\bigl(i\,Q^{(n)}_\mu(x)\bigr)\,U_\mu(x)\,,
\end{equation}
with $ Q^{(n)}_\mu(x) $ traceless Hermitian, built from staples through
\begin{equation}
\Omega^{(n)}_\mu(x)=C^{(n)}_\mu(x)\,U^\dagger_\mu(x)\,, \qquad
Q^{(n)}_\mu(x)=\frac{i}{2}\bigl(\Omega^{\dagger}-\Omega\bigr)-\frac{i}{2N}\Tr\bigl(\Omega^{\dagger}-\Omega\bigr)\,.
\end{equation}
The staple sum is
\begin{equation}
\begin{aligned}
C^{(n)}_\mu(x)=\sum_{\nu\neq\mu}\Bigl[
&\rho^{+}_{\mu\nu}(n,x)\,U_\nu(x)U_\mu(x+\hat\nu)U^\dagger_\nu(x+\hat\mu)\\
+&\rho^{-}_{\mu\nu}(n,x)\,U^\dagger_\nu(x-\hat\nu)U_\mu(x-\hat\nu)U_\nu(x-\hat\nu+\hat\mu)
\Bigr]\,.
\end{aligned}
\end{equation}
To make the Jacobian determinant tractable, we use an even–odd masking schedule, which ensures a (block-)triangular Jacobian and thus allows an efficient computation of the determinant as the product of the diagonal blocks. 

\begin{wrapfigure}{r}{0.7\textwidth}
	\centering
	\includegraphics[width=0.5\textwidth]{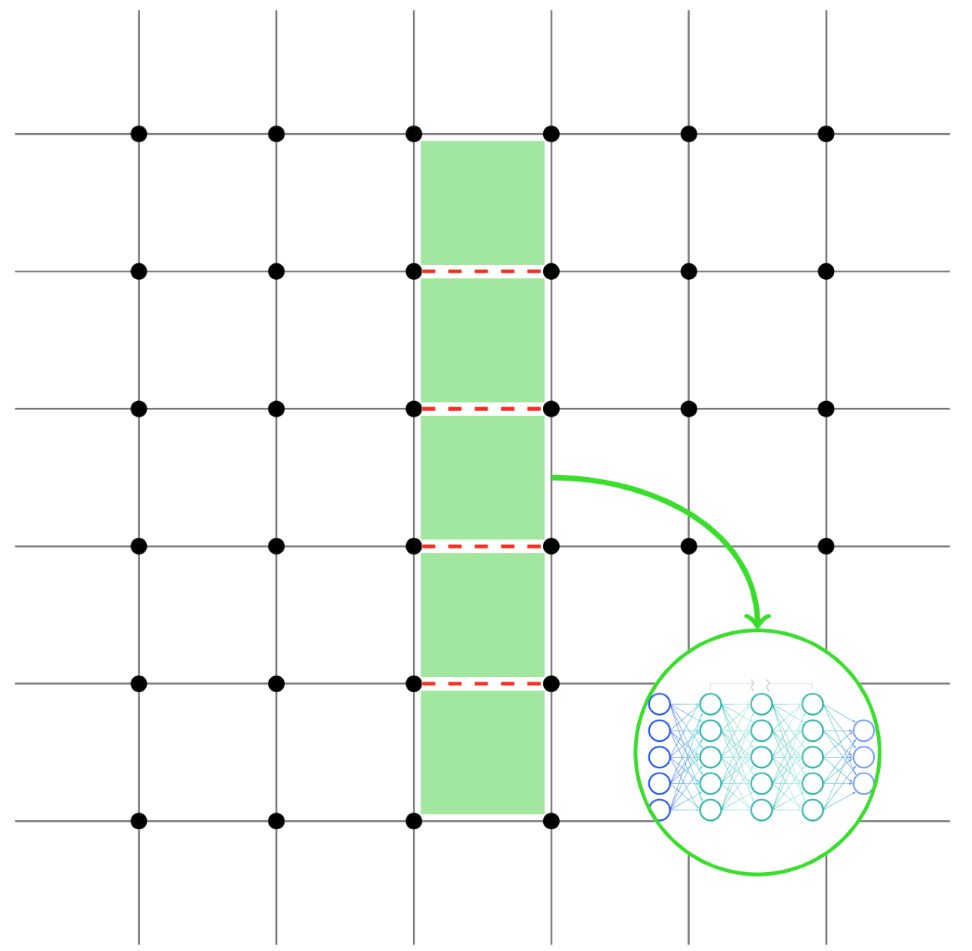}
	\caption{Schematic representation of the defect coupling layer used in the SNF. The deterministic transformation updates only the links on and in the immediate neighborhood of the defect (green strip); the red dashed links indicate the subset of boundary links whose couplings are modified during the boundary-condition evolution.}
	\label{fig:Defect}
\end{wrapfigure}

In boundary-condition evolutions the action is modified only in a localized region (the defect), thus, following the work in Ref.~\cite{Bulgarelli:2024yrz}, we restrict the deterministic layer support to the defect and its immediate neighborhood. Refer to Fig.~\ref{fig:Defect} for a schematic representation of the defect coupling layer.
This particular coupling layer reduces the number of trainable parameters and focuses the flow capacity where the mismatch to PBC is largest.
Global propagation of the defect information is still guaranteed by the interleaved stochastic gauge update, which acts on the full lattice. In this work, we use the standard heatbath plus 4 overrelaxation steps as MCMC update. 

Defect SNFs training is done by minimizing the average dissipated work,
\begin{equation}
\mathcal{L}(\rho)=\langle \Wd^{(\rho)} \rangle_{\mathrm{f}},
\end{equation}
with respect to the parameters $\rho$ of the layer. This procedure favors reversible trajectories and improves both variance and stability of reweighting. Furthermore, as shown in Ref.~\cite{Bulgarelli:2024brv,Bonanno:2025pdp}, in practice, one can train at fixed target coupling and small $ \nstep $ (e.g.\ $ \nstep=8,16 $) and observe smooth profiles as functions of $ n/\nstep $.
This enables a simple transfer procedure: group parameters into a small number of geometric classes (suggested by symmetries of the defect), and interpolate with splines a rescaled profile $ \rho_{\mathrm{class}}(n)\,\nstep $ as a function of $ n/\nstep $.
The resulting fit is then used to instantiate larger $ \nstep $ flows at negligible additional training cost. See Ref.~\cite{Bulgarelli:2024brv,Bonanno:2025pdp} for further details on the training procedure.

\section{Scaling and numerical performance}

\begin{figure}[t]
  \centering
  \includegraphics[scale=0.5]{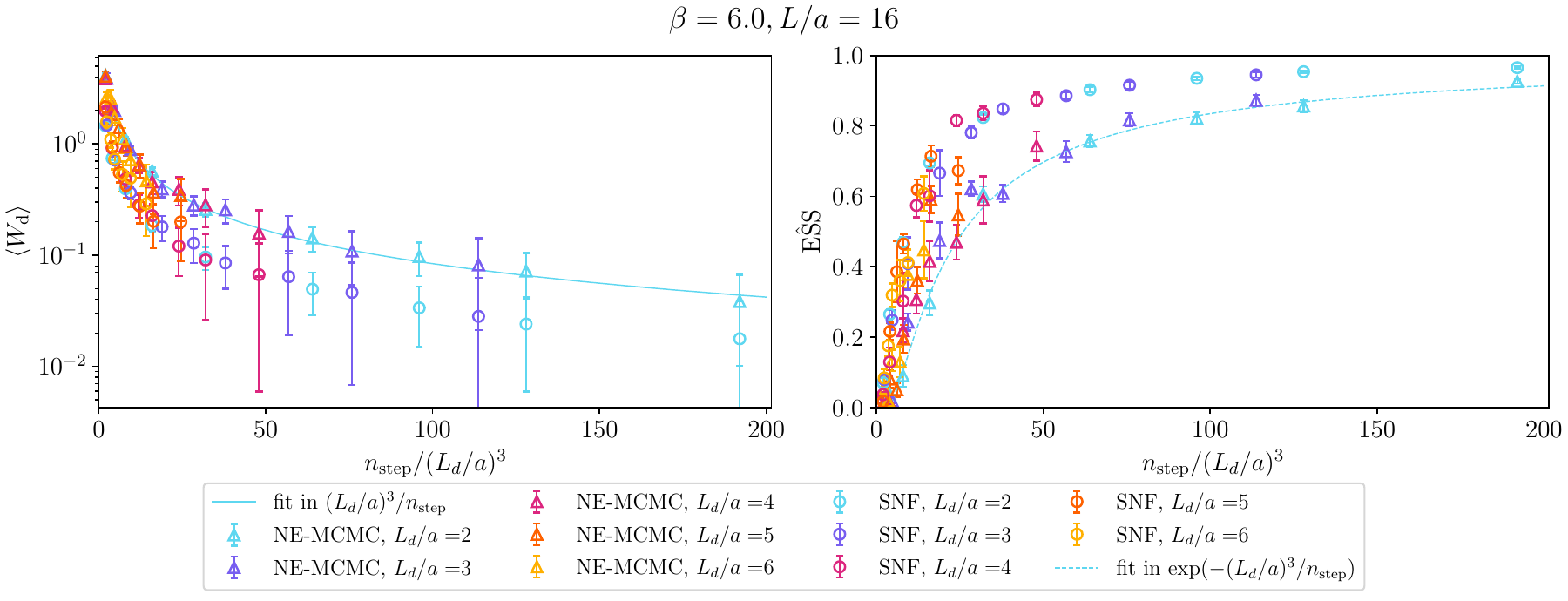}
  \caption{Comparison of the dissipated work (left) and $ \hESS $ (right) for NE-MCMC (triangles) and defect SNFs (circles) as a function of $ \nstep/\ndof $ at $\beta=6$ and $L/a=16$. Figure taken from \cite{Bonanno:2025pdp}.}
  \label{fig:dkl_ess}
\end{figure}

For boundary-condition flows, the relevant size parameter is the number of degrees of freedom affected by the defect, $ \ndof \propto (L_d/a)^3 $.
A robust empirical scaling observed for purely stochastic non-equilibrium flows is
\begin{equation}
\langle \Wd \rangle_{\mathrm{f}} \propto \frac{\ndof}{\nstep}\,,
\end{equation}
and
\begin{equation}
\hESS \approx \exp\,\Bigl(-k'\,\frac{\ndof}{\nstep}\Bigr)\,,
\end{equation}
so controlling $ \hESS $ at fixed defect size requires $ \nstep \propto \ndof $.
These scaling relations are illustrated in Fig.~\ref{fig:dkl_ess}, where we compare purely stochastic NE-MCMC and defect SNFs at $ \beta=6.0 $ on a $ L/a=16 $ lattice for several defect sizes. 
When the performance metrics are plotted against the scaling variable $ \nstep/\ndof $, data corresponding to different $ L_d/a $ values collapse onto a common curve, showing that the efficiency is primarily controlled by the ratio between the number of non-equilibrium steps and the number of degrees of freedom touched by the defect. 
Increasing $ \nstep/\ndof $ leads to smaller $ \langle \Wd\rangle_{\mathrm f} $ (left panel) and, consistently, to larger $ \hESS $ (right panel), i.e., to a better-behaved reweighting estimator. 
At fixed $ \nstep/\ndof $, defect SNFs systematically yield more reversible trajectories than NE-MCMC, resulting in a higher $ \hESS $; equivalently, for a fixed target $\hESS $ the SNF reaches the same estimator quality with fewer non-equilibrium steps, corresponding to an overall speedup of about a factor $ \sim 3 $ in this setup.

\begin{figure}[t]
  \centering
  \includegraphics[scale=0.5]{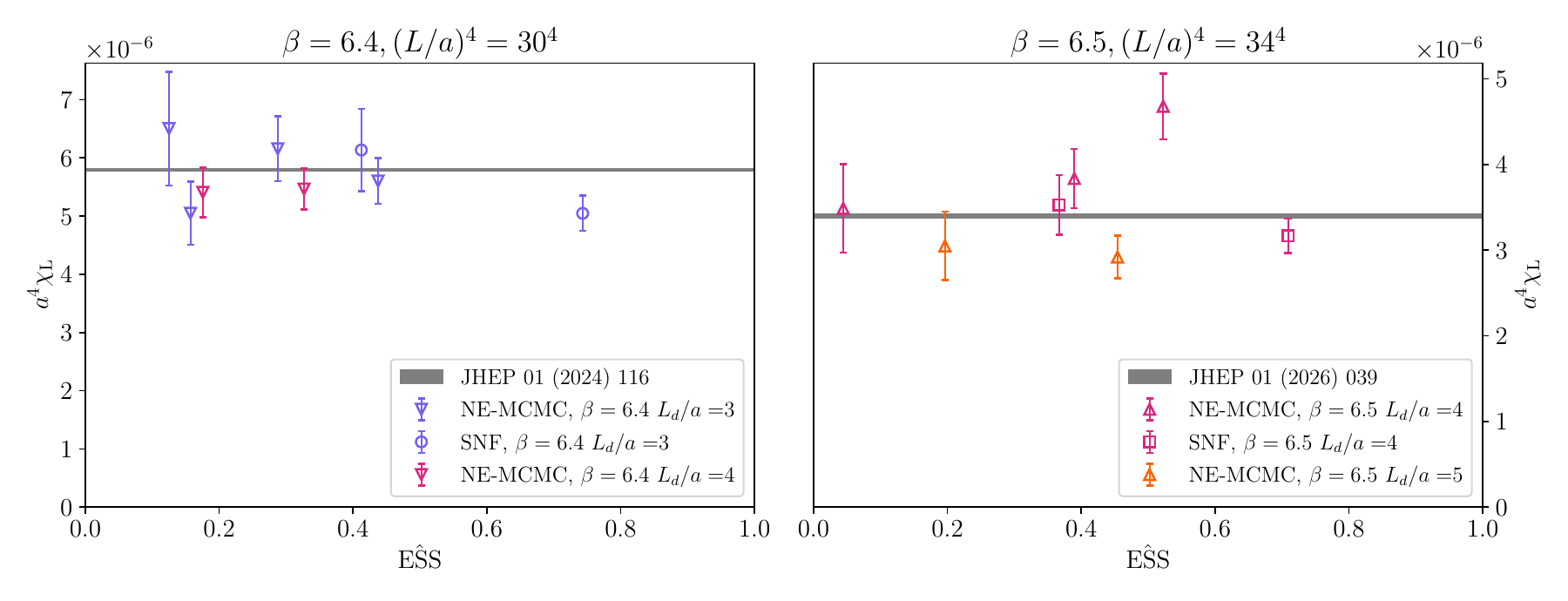}
  \caption{Topological susceptibility in lattice units $a^{4}\chi_{_{\scriptscriptstyle{\rm L}}} $ versus proxy effective sample size $ \hESS $ for boundary-condition flows at $ \beta=6.4 $ ($30^{4}$ lattice) and $\beta=6.5 $ ($34^{4}$ lattice). Horizontal bands show reference results from \cite{Bonanno:2023ple,Bonanno:2025eeb}. Figure taken from \cite{Bonanno:2025pdp}.}
  \label{fig:chi_ESS}
\end{figure}

As a physics validation of the method, Fig.~\ref{fig:chi_ESS} shows the topological susceptibility in lattice units $ a^{4}\chi_{_{\scriptscriptstyle{\rm L}}}$ obtained from the reweighted PBC ensemble as a function of the proxy effective sample size $ \hESS $. 
Results at $ \beta=6.4 $ on a $ 30^{4} $ lattice and at $ \beta=6.5 $ on a $ 34^{4} $ lattice are consistent across different flow setups and defect sizes. Moreover, our determinations agree with high-statistics reference computations, shown as horizontal bands, providing a non-trivial check that the reweigthing procedure suggested by Jarzynski equality correctly removes the defect-induced boundary artifacts and reproduces the theory with PBC.

\section{Conclusions and future outlooks}

We have introduced a defect-based Stochastic Normalizing Flow strategy to exploit the fast topological dynamics of open boundaries while recovering expectation values of the physical periodic theory through Jarzynski equality. By interleaving global non-equilibrium MCMC steps with localized, gauge-equivariant defect coupling layers, the SNF produces more reversible trajectories than purely stochastic NE-MCMC, reducing dissipation and increasing $ \hESS $ at fixed scaling variable $ \nstep/\ndof $; this yields a tangible speedup at fixed estimator quality.

Future developments include several clear directions. On the machine learning side, richer gauge-equivariant layers and multiscale architectures~\cite{Bauer:2024byr,Singha:2025lsd,Ihssen:2025ybn} can improve the gain of SNFs on NE-MCMC. On the algorithmic side, improving the non-equilibrium evolution itself is a natural direction, in particular by optimizing the schedule for the boundary-interpolation parameter (beyond a linear protocol) to reduce the dissipated work at fixed computational cost. Finally, the same framework can be extended to more challenging settings (including dynamical-fermion simulations), with the potential to enable controlled topology sampling closer to the continuum limit.

\section*{Acknowledgments}
\noindent
We thank M.~Caselle, G.~Kanwar and M.~Panero for insightful and helpful discussions. C.~B.~acknowledges support by the Spanish Research Agency (Agencia Estatal de Investigación) through the grant IFT Centro de Excelencia Severo Ochoa CEX2020-001007-S and, partially, by grant PID2021-127526NB-I00, both funded by MCIN/AEI/10.13039/ 501100011033. A.~B., A.~N., D.~P.~and L.~V. acknowledge support by the Simons Foundation grant 994300 (Simons Collaboration on Confinement and QCD Strings). A.~B.~was funded by the Deutsche Forschungsgemeinschaft (DFG, German Research Foundation) as part of the CRC 1639 NuMeriQS – project no. 511713970 and under Germany’s Excellence Strategy – Cluster of Excellence Matter and Light for Quantum Computing (ML4Q) EXC 2004/1 – 390534769. A.~N.~acknowledges support from the European Union - Next Generation EU, Mission 4 Component 1, CUP D53D23002970006, under the Italian PRIN “Progetti di Ricerca di Rilevante Interesse Nazionale – Bando 2022” prot. 2022ZTPK4E. A.~B., A.~N., D.~P.~and L.~V.~acknowledge support from the SFT Scientific Initiative of INFN. The work of D.~V.~is supported by STFC under Consolidated Grant No.~ST/X000680/1. 
We acknowledge EuroHPC Joint Undertaking for awarding the project ID EHPC-DEV-2024D11-010 access to the \texttt{LEONARDO} Supercomputer hosted by the Consorzio Interuniversitario per il Calcolo Automatico dell'Italia Nord Orientale (CINECA), Italy. This work was partially carried out using the \href{https://www.plymouth.ac.uk/about-us/university-structure/faculties/science-engineering/hpc}{computational facilities of the ``Lovelace'' High Performance Computing Centre, University of Plymouth}.

\bibliographystyle{JHEP}
\bibliography{biblio}

\end{document}